\documentclass[fleqn,10pt]{wlscirep}
\usepackage{graphicx}
\usepackage{xcolor}
\usepackage{amsmath, amsfonts, amssymb}
\DeclareGraphicsExtensions{.jpg,.png,.eps,.svg,.pdf}
\usepackage{epstopdf}
\usepackage{subfigure}
\usepackage{enumerate}
\usepackage{multirow}
\usepackage{float}
\usepackage{lineno}
\title{Origin of multiple band gap values in single width nanoribbons}

\author[1]{Deepika}
\author[2]{Shailesh Kumar}
\author[3]{Alok Shukla}
\author[1,*]{Rakesh Kumar}
\affil[1]{Department of Physics, Indian Institute of Technology Ropar, Rupnagar-140001, India}
\affil[2]{School of Chemistry, Physics and Mechanical Engineering, Queensland University of Technology, Brisbane, Queensland 4000, Australia}
\affil[3]{Department of Physics, Indian Institute of Technology Bombay, Powai, Mumbai-400076, India}

\affil[*]{corresponding:rakesh@iitrpr.ac.in}

\begin{abstract}
Deterministic band gap in quasi-one-dimensional nanoribbons is prerequisite for their integrated functionalities in high-performance molecular-electronics based devices. However, multiple band gap values commonly observed in the same width of graphene nanoribbons fabricated in same slot of the experiments remains unresolved, and raise a critical concern over scalable production of pristine and/or hetero-structure nanoribbons with deterministic properties and functionalities for plethora of applications. Here, we show that a modification in the depth of potential wells in the periodic direction of a supercell on relative shifting of passivating atoms at the edges is the origin of multiple band gap values for the same width of nanoribbons in a crystallographic orientation, although they carry practically the same ground state energy. The results are similar when calculations are extended from planar graphene to buckled silicene nanoribbons. Thus, the findings facilitate tuning of the electronic properties of quasi-one-dimensional materials such as bio-molecular chains, organic and inorganic nanoribbons by performing edge engineering.\\
 
\end{abstract}
\begin{document}

\flushbottom
\maketitle
%
%
%

\section*{Introduction}

Quasi-one-dimensional nanoribbons in their sub-10 nm width are focus of current research interest amongst low dimensional materials due to their exceptional promises to enable new functionalities, and improve the performance of devices down to molecular level \cite{van_der_lit_suppression_2013, welte_highly_2010, zhong_helical_2014, cai_atomically_2010, gracia-espino_fabrication_2015}. Nevertheless, band gap of    graphene nanoribbons (GNRs) \cite{wallace_band_1947, nakada_edge_1996, son_energy_2006} is critical in addition to its other interesting electronic properties such as quasi-relativistic behavior of charge carriers \cite{novoselov_two-dimensional_2005}, the highest thermal conductivity \cite{balandin_superior_2008}, and the highest mobility \cite{bolotin_temperature-dependent_2008} of charge carriers at room temperature for such high performance applications \cite{novoselov_roadmap_2012, geim_rise_2007}. The common experimental observations of multiple band gaps due to different passivation patterns at the edges in the same width of graphene nanoribbons (GNRs) have been neglected so far, since the focus of research on GNRs has been to understand the experimental observation of non-zero band gap values in both the crystallographic orientations. Multiple band gaps were observed in the same width of GNRs fabricated even in the first experiment performed on GNRs in 2007 \cite{han_energy_2007}. Later, several experimental papers also reported multiple band gap values corresponding to different passivation patterns at the edges in the same width of GNRs and other quasi-one dimensional materials such as silicon nanowires fabricated in a crystallographic orientation \cite{li_chemically_2008, zhu_physicochemical_2013, tseng_diluted_2009, molitor_energy_2010, schwierz_graphene_2010, mohammad_understanding_2014}. Since 2007, hundreds of theoretical papers have been published on GNRs focused only on how to explain non-zero band gap in both the crystallographic orientations of GNRs based on different edge passivating  patterns with different type of edge passivating elements.  \cite{sols_coulomb_2007, liao_charge_2012, hod_edge_2007, cruz-silva_edgeedge_2013, lee_electronic_2009, ramasubramaniam_electronic_2010, liu_toward_2015, deepika_edge_2015}, the multiple band gap values in the same width of nanoribbons remains unresolved. Therefore, in this work, we for the first time propose a resolution for the physical origin of multiple band gap values in single width nanoribbons, which would resolve the critical concern over the scalable production of pristine and/or hetero-structure nanoribbons with deterministic properties for plethora of applications.

For theoretical investigation of the multiple band gap values in the same width of nanoribbons, we consider supercells in a crystallographic orientation with the same number of atoms. The supercells are different only in terms of arrangement of atoms at the edges with respect to each other. We have chosen oxygen as passivating atoms at the edges of nanoribbons because oxygen plasma is commonly used in fabrication of GNRs. The considered supercell edge configurations of GNRs are energetically favorable with respect to other possible edge configurations having the same number of atoms in a crystallographic orientation (\textit{cf.} supplementary information S1). Band structure calculations show an appreciable change in the band gap values in the same width of nanoribbons, however their ground state energy are practically same. Based on theoretical analysis, it is found that the modification in the Columbic potential profiles in the periodic direction is the origin of multiple band gap values in the fabricated GNRs. The results are verified when calculations are extended from planar graphene nanoribbons to buckled silicene nanoribbons (SiNRs), GNRs passivated with multiple functional groups at the edges, and rough edged GNRs. Thus, the findings facilitate in tuning the electronic properties of quasi-one-dimensional materials such as bio-molecular chains, organic and inorganic nanoribbons by edge engineering, which improve the performance of devices down to molecular level for their wide applications  \cite{lemme_graphene_2007, pumera_graphene_2011, shukla_graphene_2009, nair_fine_2008, yoon_dissipative_2012}.

\section*{Results and Discussions}

\begin{figure}[!ht]
 \centering
\includegraphics[width=\linewidth]{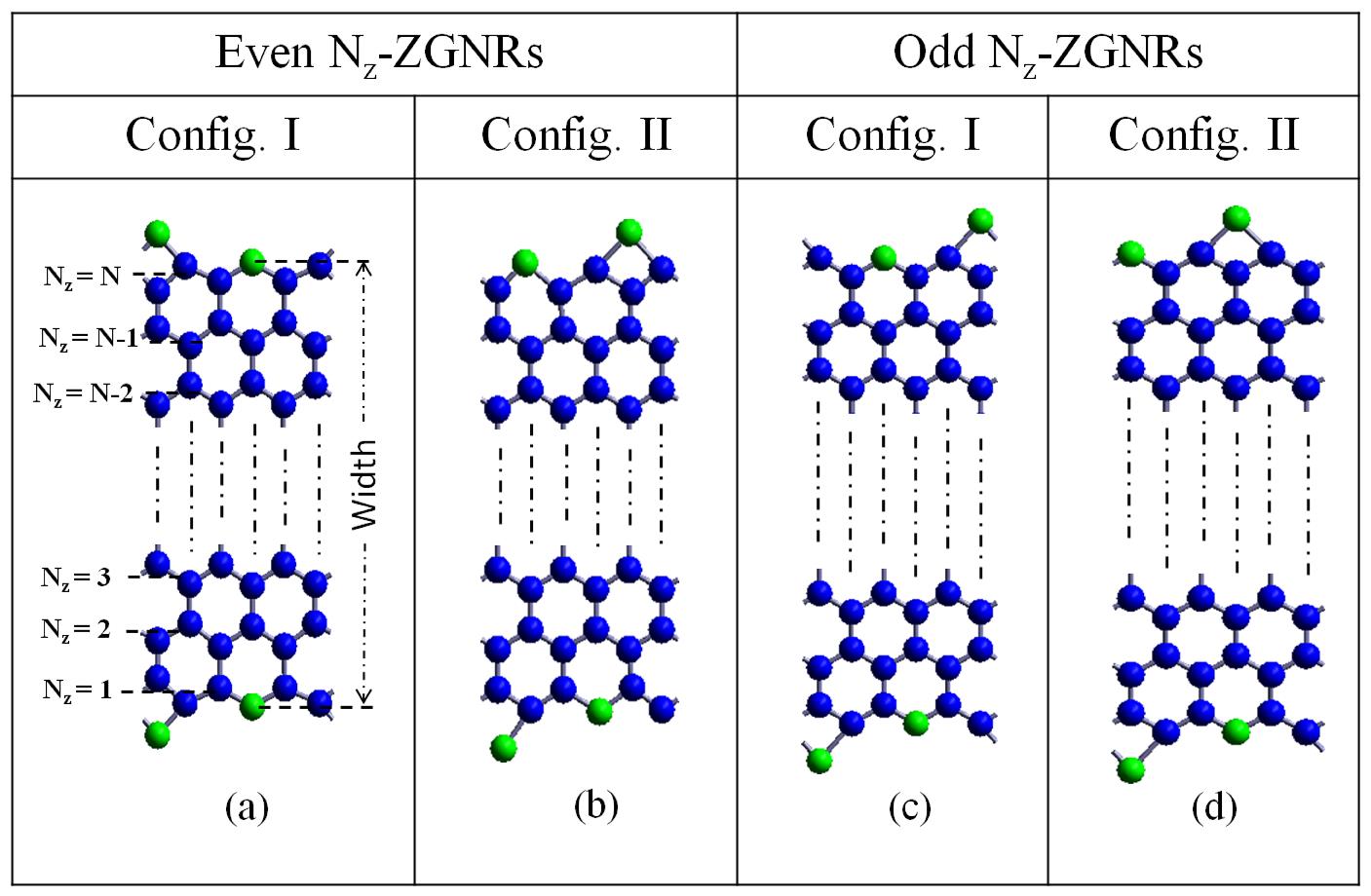}
\caption{Two possible edge configurations for ZGNRs supercell, (a) and (b) represents configurations corresponding to even N$_z$-ZGNRs, while (c) and (d) corresponds to odd N$_z$-ZGNRs. Blue and green spheres represent carbon and oxygen atoms, respectively. Note: In both configurations, the bottom edge remains the same, only the relative arrangement of atoms on the top edge is different.} 
\label{configzgnrs}
\end{figure}

\begin{figure}[!t]
\includegraphics[width=\linewidth, keepaspectratio]{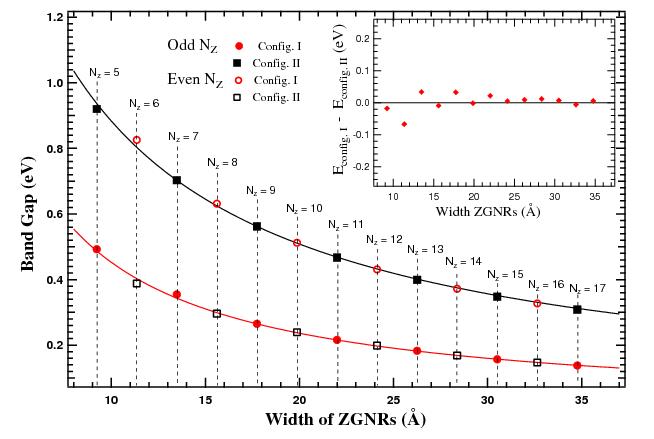}
\caption{Energy band gap values as a function of width corresponding to configurations I and II of ZGNRs. Solid curves shows fitting to the band gap values using scaling formula $\Delta E$ = $\frac{\alpha}{(w+w^{`})}$ for GNRs, where $\Delta E$ is the band gap (eV), $\alpha$ (eV.\AA$^{-1}$) and $w^{`}$(\AA{}) are the scaling factors. Dotted vertical lines are drawn to represent two band gap values for the same width of ZGNRs. Inset figure shows the difference of the ground state energy values for config. I and config. II as a function of width for ZGNRs.} 
\label{bandgapzgnrs}
\vspace{-0.1cm}
\end{figure}

\noindent To investigate the origin of multiple band gap values for the same width of nanoribbons in a crystallographic orientation, we consider GNRs with two edge configurations (config. I and config. II). For ZGNRs, two edge configurations (I and II) for supercells corresponding to even N$_z$ and odd N$_z$ are shown in Fig. \ref{configzgnrs}. Band structures are calculated from  $\Gamma$ (k = 0) to X-point (k = $\pi$) upto a maximum width of $\approx$ 35 \AA{} (N$_z$ = 17). Typical band structures for N$_z$= 5 and 6-ZGNRs of config. I and config. II are shown in supplementary information S2. Non-zero direct band gaps are observed at $\Gamma$ point for both the configurations. An appreciable difference in the band gap values are observed for the same width of ZGNRs (Fig. \ref{bandgapzgnrs}). The difference in the band gap values decreases with increase in the width of ZGNRs. It is to be noted that even after having significant change in the band gap values for the same width of ZGNRs, their ground state energies remain practically the same (inset of Fig. \ref{bandgapzgnrs}). The maximum difference in the ground state energy is $\approx$ 0.06(6) eV for N$_z$ = 6, which further decreases with the width. Electrostatic edge-edge interactions is one possible reason behind the small difference in the ground state energies of both the  configurations. To explore this possibility further, we placed discrete electronic charges on the atomic sites at the edges, and calculated the difference of the electrostatic energy for the two configurations. This energy difference, although a bit larger, is of the same order of magnitude as compared to the difference between the ground state energies of both the configurations for the same width of nanoribbons. This result suggests the Columbic nature of the edge-edge interactions, and relates the change in the ground state energy with the modification in the potential energy as a consequence of change in the arrangement of atoms at the edges. \\

\noindent Since one dimensional periodic potential is the limiting case for the periodic potential of a quasi-one-dimensional nanoribbons, therefore an average of the local potentials along the width of the nanoribbon projected on periodic direction may be considered to explain the difference in the band gap values. In order to investigate it, we plotted potential profiles (average of local potentials in the periodic direction) of the ZGNRs supercells, which looks similar to that of Kronig-Penney (KP) potential wells. Typical potential profiles corresponding to config. I and config. II for both N$_z$ = 5 and 6 are shown in Fig. 3.  The number of potential wells in a supercell along the periodic direction is equal to that of the atomic YZ planes, and a change in potential profiles corresponding to config. I and config. II for the same width of ZGNRs is observed [Fig. 3(a), 3(b)]. From KP model, for the same width of potential wells, the band gap is proportional to the depth of well. Therefore, the depth of the deepest potential well corresponding to global minima is compared for both the configurations. The configuration with the deepest global minima amongst the potential wells is found to have higher band gap value in agreement with the theory of KP model. The normalized potential depth of the deepest global minima (w.r.t. N$_z$) is plotted as a function of width for the config. I and config. II for ZGNRs (Fig. \ref{potentialdepth}). It is to be noted that as N$_z$ is changed to N$_{z+1}$, the deepest potential well switches from one configuration to another configuration, and accordingly higher band gap value also switches to another configuration (Fig. \ref{bandgapzgnrs} and \ref{potentialdepth}). This explains a change in the depth of the deepest global minima in potential profile of a supercell, resulting from a change in the arrangement of atoms at the edges, to be the origin of multiple band gap values for the same width of ZGNRs. The difference in the band gap values  decreases similar to that of the average normalized potentials for higher widths of ZGNRs (Fig. \ref{potentialdepth}), which indicates the decreasing edge effects on the average of local potentials in higher widths of ZGNRs. Therefore, the effect of a change in the potential profile on the band gap is primarily a consequence of modification in the electrostatic interactions among the charges at the edges of nanoribbons. \\

\noindent For investigating the effect of potential profiles on band gap in armchair GNRs, two edge configurations for N$_a$-AGNRs are considered similar to that of ZGNRs. Two edge configurations for AGNRs supercell of odd N$_a$ and even N$_a$ are shown in Fig. \ref{configagnrs}. Band structures are calculated up to a maximum width of $\approx$ 23 \AA{} (N$_a$ = 20). Typical band structures for N$_a$= 5 and 6-AGNRs of config. I and config. II are shown in supplementary information S3. Direct band gap is observed at $\Gamma$ (k = 0) point for both the configurations. It is to be noted that the multiple band gap values are observed only for odd N$_a$-AGNRs, while the same band gap values are observed for even N$_a$-AGNRs. The difference in band gap values between both the configurations of odd N$_a$-AGNRs decreases with the increase in the width. The ground state energy corresponding to both the configurations of AGNRs are nearly same (negligible change in the third decimal place) for supercells of the same width, except for N$_a$ = 5 and 7.  The exceptional behavior for N$_a$ = 5 and 7-AGNRs has been discussed in supplementary S4 \\

\noindent Similar to ZGNRs, potential profiles of AGNRs are plotted for both the configurations in periodic direction of the supercells. For odd N$_a$-AGNRs, the edge configuration with the deepest potential well at global minima corresponds to the higher band gap value except for N$_a$ = 5 and 7. For even N$_a$-AGNRs, the potential profiles of both the configurations superpose on each other on relative shifting in the periodic direction as shown in Fig. \ref{potentialagnrs}. Therefore, both the configurations of even N$_a$-AGNRs correspond to the same band gap values, while different band gap values are observed for odd N$_a$-AGNRs. The difference in the band gap values of both the configurations for odd N$_a$-AGNRs becomes negligible for N$_a$ = 15 onwards. Nevertheless, a small but nearly constant difference between the deepest potential wells at global minima of the configurations for higher widths is observed possibly as a consequence of inherent potential associated with the configurations. \\

\indent In brief, multiple band gap values for the same width of GNRs in a crystallographic orientation depends upon the relative arrangement of atoms at the edges in  the supercells. On changing the arrangement of atoms at the edges, band gap is changed only for an asymmetrical modification in potential profiles of  the supercells. Band gap is found to be higher for a configuration with the deepest potential well at the global minima amongst the same width of GNRs. How fast the difference in the band gap values between both the configurations of GNRs would decrease, depends upon arrangement of atoms at the edges in a supercell.

\indent To generalize the concept for multiple band gaps in nanoribbons, we apply configurational change at the edges to the same width of buckled silicene nanoribbons (\textit{cf.} supplementary information S5) as it is done for GNRs. In the case of silicene, oxygen passivated zigzag configuration similar to ZGNRs is theoretically not possible due to higher value of Si-Si bond lengths $\approx$ 2.23 \AA{} (lattice constant for silicene lattice $\approx$ 3.826 \AA{}) in comparison to Si-O bond length of $\approx$ 1.538 \AA{}. Therefore, the studies are carried only for armchair silicene nanoribbons (ASiNRs). The results are similar to GNRs. Multiple band gap values are observed on changing the arrangement of atoms at the edges for the same width of odd N$_a$-ASiNRs, and no difference is observed for even N$_a$-ASiNRs. Higher band gap value corresponds to the configuration with the deepest potential well at global minima except for N$_a$ = 5 and 7. The exceptional behavior for N$_a$ = 5 and 7 is found to be the same as that for N$_a$ = 5 and 7-AGNRs. The difference in the band gap values between both configurations of SiNRs decreases with width similar to GNRs, which reflects the decreasing edge-contributions to the potential profiles in higher width of nanoribbons.

\indent In order to verify the concept of multiple band gaps for nanoribbons passivated with different types of atoms at the edges,  such as bio-molecular chains passivated with different group of elements at the edges, we consider GNRs passivated with two different types of atoms such as hydrogen and oxygen. Similar to the above findings, multiple band gap values are observed for the same width of GNRs with practically the same ground state energy (\textit{cf.} supplementary information S6). The effect of change in the arrangement of atoms at the edges is also observed on band gap in the rough edged GNRs passivated with oxygen atoms; a significant change in the band gap values are observed for the same width of rough edged GNRs with practically the same ground state energy (\textit{cf.} supplementary information S7). The finding is relevant for explanation of the experimental observations of multiple band gap values for the same width of GNRs fabricated in a crystallographic orientation using oxygen plasma etching process. \\

\section*{Conclusions}

On the basis of our first-principles calculations, it is concluded that

\begin{itemize}
\item Origin of multiple band gap values in nanoribbons of the same width, same crystallographic orientation, and the same number of the atoms in the supercells is a consequence of modification in the potential profile in the periodic direction, although they carries practically the same ground state energy.
\item The modification in the depth of the potential well is primarily a consequence of change in the electrostatic interactions among the charges at the edges of nanoribbons, which arise from a change in the arrangement of atoms at the edges. 
\item Asymmetrical modification in the potential profiles results into a significant change in the band gap value of a nanoribbon, while symmetrical modification leads to the same band gap values.
\item The configuration with the deepest potential well at the global minima in the potential profile of a supercell corresponds to the highest band gap value of a nanoribbon except for the supercells having modification in the bond lengths on a change in the arrangement of atoms at the edges.  
\item The difference in the band gap values of both the configurations decreases  with increase in width of nanoribbons, which indicates decreasing edge effects on the potential profiles,  therefore would converge at higher width. How fast it would converge depends typically on arrangement of atoms at the edges.
\end{itemize}

\indent In remark, multiple band gap values are possible by a change in the relative arrangement of atoms at the edges in the same width of nanoribbons. The concept can be applied to understand the experimental observations of multiple band gap values in the same width of nanoribbons, and also the physical origin of the differential behavior in bio-molecules such as protein and carbohydrates, where only the relative position of the molecular functional group changes at the edges. These findings are critical for edge engineering of quasi-one-dimensional materials and bio-molecules. With deterministic control at the edges, the properties of quasi-one-dimensional materials and bio-molecules can be tuned for designing new biomaterials \cite{rouge_biomolecules_2011}, molecular level solution to the problems related to environment, food industry, biotechnology and medicines.

\section*{Methods}
 
\noindent We performed first-principles band structure calculations for nanoribbons based on density functional theory (DFT) using Vienna \textit{ab-initio} simulation package (VASP) \cite{kresse_efficiency_1996}. Generalized gradient approximations are used as exchange-correlation functional with electron-ion interactions in projected augmented wave formalism. The cutoff energy of 400 eV and 800 eV is used for graphene and silicene, respectively with a vacuum layer of  $\approx$ 10 \AA{}. The relaxation of the system is performed until  the force experienced by each atom is $\leq$ 0.001 eV \AA{}$^{-1}$. The Monkhorst-Pack k-space mesh of 25x1x1 is used for the k-space sampling. All the calculations for nanoribbons are performed for nonmagnetic ground state, since the ground state is found to be nonmagnetic upon introducing spin-polarization. Number of zigzag chains or dimer lines along normal to the periodic direction is used to represent the width of the nanoribbons, and higher width of nanoribbons 
is obtained by adding pure carbon chains or dimer lines. The number of zigzag chains (N$_z$) for zigzag graphene nanoribbons (ZGNRs) corresponds to N$_z$-ZGNRs and number of dimer lines (N$_a$) for armchair graphene nanoribbons (AGNRs) as N$_a$-AGNRs; and the similar nomenclature is applied to SiNRs.


\section*{Acknowledgments}

We gratefully acknowledge IIT Ropar for providing the High Performance Supercomputing facility and CDAC-Pune for the PARAM YUVA II Supercomputing facility. R. Kumar and Deepika acknowledges P. Wadhwa for her support in data collection during preparation of the manuscript. 

\section*{Author contributions statement}

R.K. conceived and supervised the study. Deepika, S.K., A.S. and R.K. developed the model for study. Deepika carried out the computations. All authors discussed and analyzed the results and co-wrote the paper. All authors reviewed the manuscript. 

\section*{Additional information}
\textbf{Supplementary information} accompanies this paper at  \textbf{http://www.nature.com/srep}. \\ 
\textbf{Competing financial interests}: The authors declare no competing financial interests. \\

\section*{Figure Legends}

\begin{figure}[!ht]
\centering
\includegraphics[scale=0.30]{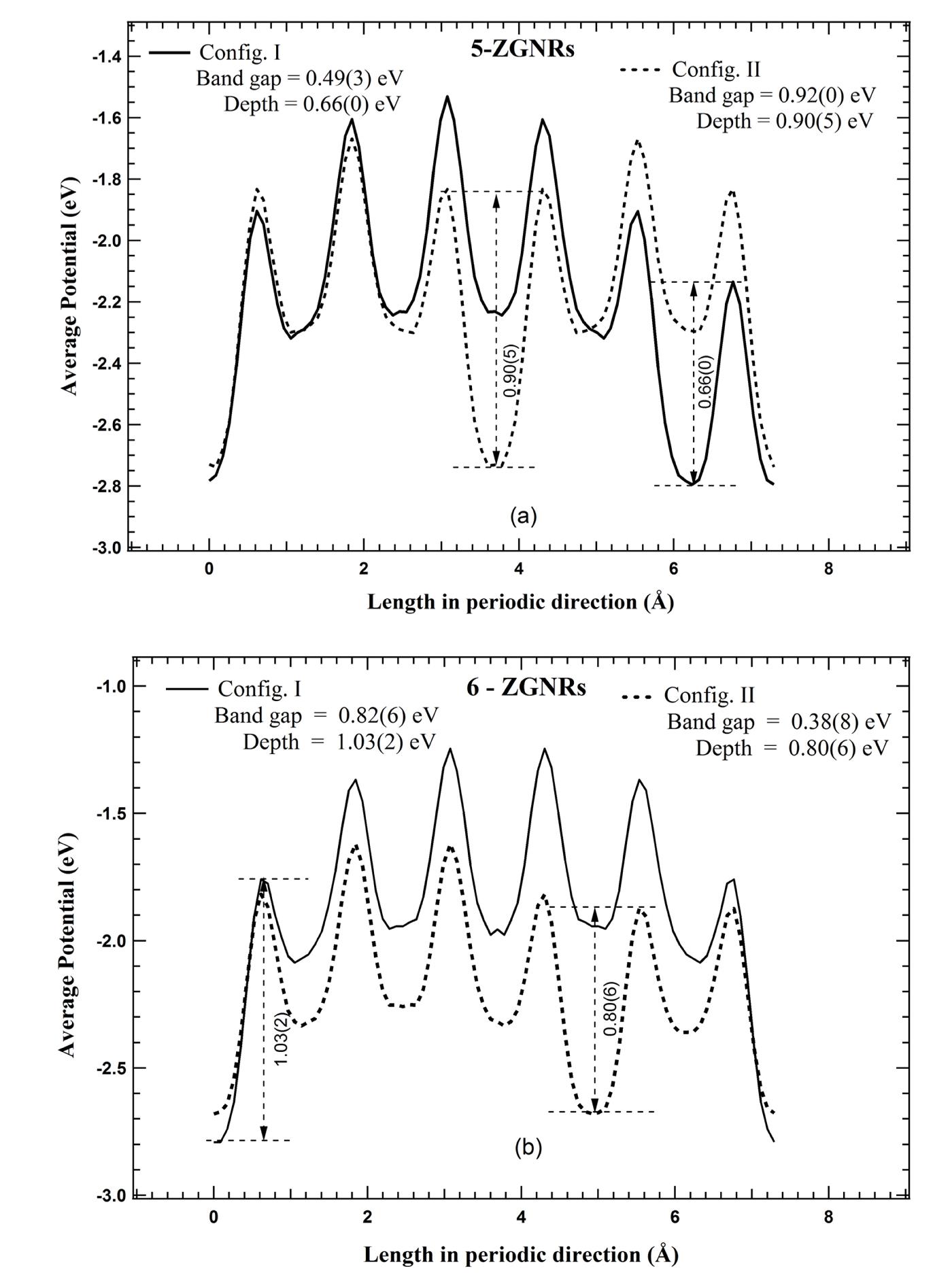}
\caption{Average of the local potentials of YZ  atomic plane plotted in periodic direction of the supercell for (a) 5-ZGNRs and (b) 6-ZGNRs. Note: Switching of the deepest potential well and band gap values from config. II (N$_z$ = 5) to config. I (N$_z$ = 6).}
\end{figure}

 \begin{figure}[!ht]
\includegraphics[width=\linewidth]{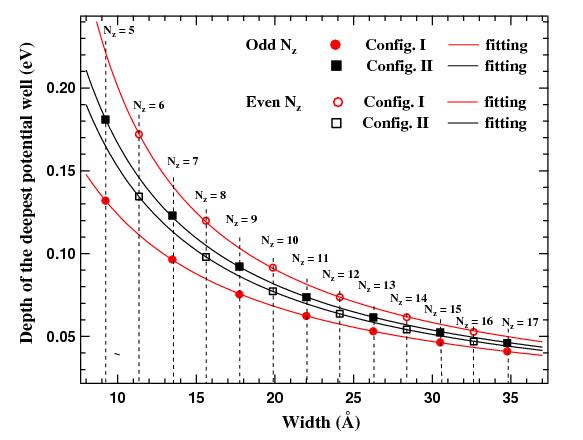}
\caption{The normalized depth of the deepest potential well (w.r.t. N$_z$) plotted as a function of width for config. I and config. II of ZGNRs. Solid curves correspond fitting to the scaling formula same as that for band gap values (see caption of Fig. 2.)} 
\label{potentialdepth}
\vspace{-0.1cm}
\end{figure}

\begin{figure}[!t]
\includegraphics[width=\linewidth]{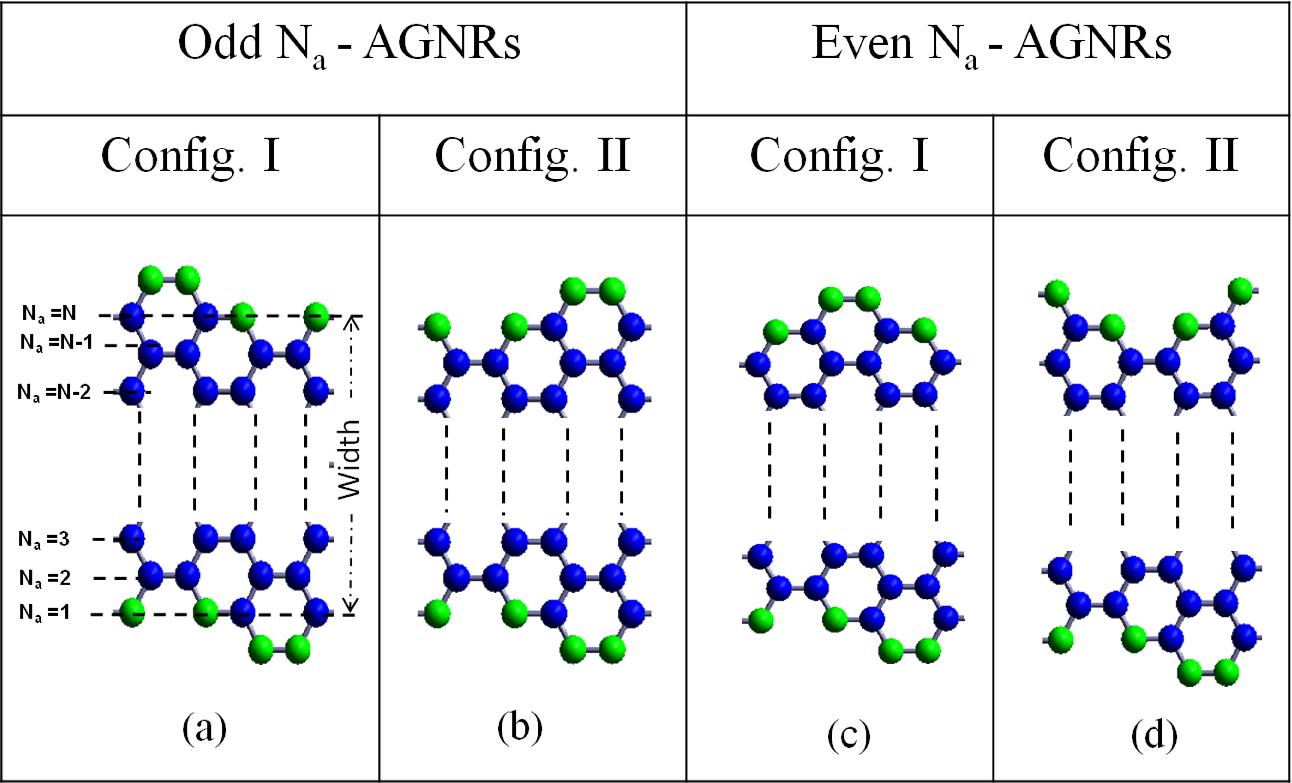}
\caption{Two possible edge configurations for AGNRs supercell corresponding to odd N$_a$ [(a),(b)] and even N$_a$ [(c), (d)]. Blue and green spheres represent carbon and oxygen atoms, respectively. Note: Change in relative arrangement of atoms at edges in config. I and config II.} 
\label{configagnrs}
\end{figure} 

\begin{figure}[!t]
\centering
\includegraphics[width=\linewidth]{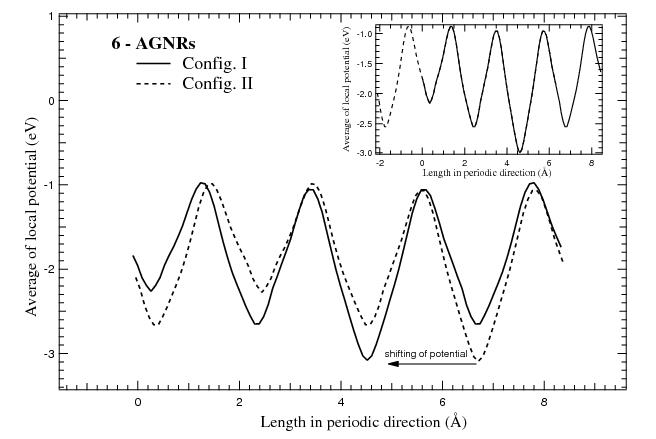}
\caption{The average of local potentials plotted along the periodic direction of 6-AGNRs supercell. Inset shows the superposition of the potential profiles on a relative shifting of config. II w.r.t. config. I.} 
\label{potentialagnrs}
\vspace{-0.1cm}
\end{figure}

\end{document}